\documentclass[]{article}
\usepackage{lmodern}
\usepackage{amssymb,amsmath}
\usepackage{ifxetex,ifluatex}
\usepackage{fixltx2e} 
\ifnum 0\ifxetex 1\fi\ifluatex 1\fi=0 
  \usepackage[T1]{fontenc}
  \usepackage[utf8]{inputenc}
\else 
  \ifxetex
    \usepackage{mathspec}
  \else
    \usepackage{fontspec}
  \fi
  \defaultfontfeatures{Ligatures=TeX,Scale=MatchLowercase}
\fi
\IfFileExists{upquote.sty}{\usepackage{upquote}}{}
\IfFileExists{microtype.sty}{%
\usepackage{microtype}
\UseMicrotypeSet[protrusion]{basicmath} 
}{}
\usepackage[margin=1in]{geometry}
\usepackage{hyperref}
\PassOptionsToPackage{usenames,dvipsnames}{color} 
\hypersetup{unicode=true,
            pdftitle={Merits and Limits: Applying open data to monitor open access publications in bibliometric databases},
            pdfauthor={Aliakbar Akbaritabar; Stephan Stahlschmidt},
            colorlinks=true,
            linkcolor=blue,
            citecolor=blue,
            urlcolor=blue,
            breaklinks=true}
\usepackage{longtable,booktabs}
\usepackage{graphicx,grffile}
\makeatletter
\def\maxwidth{\ifdim\Gin@nat@width>\linewidth\linewidth\else\Gin@nat@width\fi}
\def\maxheight{\ifdim\Gin@nat@height>\textheight\textheight\else\Gin@nat@height\fi}
\makeatother
\setkeys{Gin}{width=\maxwidth,height=\maxheight,keepaspectratio}
\IfFileExists{parskip.sty}{%
\usepackage{parskip}
}{
\setlength{\parindent}{0pt}
\setlength{\parskip}{6pt plus 2pt minus 1pt}
}
\providecommand{\tightlist}{%
  \setlength{\itemsep}{0pt}\setlength{\parskip}{0pt}}
\ifx\paragraph\undefined\else
\let\oldparagraph\paragraph
\renewcommand{\paragraph}[1]{\oldparagraph{#1}\mbox{}}
\fi
\ifx\subparagraph\undefined\else
\let\oldsubparagraph\subparagraph
\renewcommand{\subparagraph}[1]{\oldsubparagraph{#1}\mbox{}}
\fi

\let\rmarkdownfootnote\footnote%
\def\footnote{\protect\rmarkdownfootnote}

\usepackage{titling}

  \title{Merits and Limits: Applying open data to monitor open access
publications in bibliometric databases}
    \pretitle{\vspace{\droptitle}\centering\huge}
  \posttitle{\par}
    \author{Aliakbar Akbaritabar\footnote{German Centre for Higher Education
  Research and Science Studies (DZHW), Schützenstr. 6a, Berlin, 10117
  (Germany);
  \href{mailto:akbaritabar@dzhw.eu}{\nolinkurl{akbaritabar@dzhw.eu}};
  (corresponding author)} \\ Stephan Stahlschmidt\footnote{German Centre for Higher Education
  Research and Science Studies (DZHW);
  \href{mailto:stahlschmidt@dzhw.eu}{\nolinkurl{stahlschmidt@dzhw.eu}}}}
    \preauthor{\centering\large\emph}
  \postauthor{\par}
    \date{}
    \predate{}\postdate{}
  
\usepackage{booktabs}
\usepackage{longtable}
\usepackage{array}
\usepackage{multirow}
\usepackage[table]{xcolor}
\usepackage{wrapfig}
\usepackage{float}
\usepackage{colortbl}
\usepackage{pdflscape}
\usepackage{tabu}
\usepackage{threeparttable}
\usepackage{threeparttablex}
\usepackage[normalem]{ulem}
\usepackage{makecell}

\begin{document}
\maketitle
\begin{abstract}
Identifying and monitoring Open Access (OA) publications might seem a
trivial task while practical efforts prove otherwise. Contradictory
information arise often depending on metadata employed. We strive to
assign OA status to publications in Web of Science (WOS) and Scopus
while complementing it with different sources of OA information to
resolve contradicting cases. We linked publications from WOS and Scopus
via DOIs and ISSNs to Unpaywall, Crossref, DOAJ and ROAD. Only about
50\% of articles and reviews from WOS and Scopus could be matched via a
DOI to Unpaywall. Matching with Crossref brought 56 distinct licences,
which define in many cases the legally binding access status of
publications. But only 44\% of publications hold only a single licence
on Crossref, while more than 50\% have no licence information submitted
to Crossref. Contrasting OA information from Crossref licences with
Unpaywall we found contradictory cases overall amounting to more than
25\%, which might be partially explained by (ex-)including green OA. A
further manual check found about 17\% of OA publications that are
\emph{not accessible} and 15\% non-OA publications that are
\emph{accessible} through publishers' websites. These preliminary
results suggest that identification of OA state of publications denotes
a difficult and currently unfulfilled task.
\end{abstract}


\section{Introduction}\label{intro}

Open access (henceforth OA) in scholarly communication describes
unrestricted access to published peer-reviewed documents written by and
addressed to researchers. These documents have traditionally been
disseminated via publications in scientific journals, which charge for
access to the respective content. Stimulated by a call for greater
openness and transparency in general (``open science''), the OA movement
has nowadays been accepted as one, though not the only, alternative for
the dissemination of scholarly documents. Even publishers seem to
embrace this new model as providing a suitable infrastructure while at
the same time securing their own economic interests.

This inter-mixture of interests has resulted not only in one, but
several forms of OA publications such as \emph{Gold}, \emph{Hidden
Gold}, \emph{Hybrid}, \emph{Green}, \emph{Delayed}, \emph{Bronze} and
\emph{Black} which are mainly based on right to access and pay to
publish models depending on venues where the OA publication is
accessible.

Due to the individual ascription of single publications to one or
several of these categories and the decentralized structure of the
scientific publishing market with a variety of diverse publishers, the
identification of OA is less trivial than it might seem. Even large
bibliometric data provider rely on external information to provide
information on OA\footnote{\url{https://clarivate.com/blog/easing-access-to-open-access-clarivate-analytics-partners-with-impactstory/}}
and most large scale undertakings by the scientometric community to
obtain reliable information on OA prevalence rely on the use of web
crawlers (Archambault et al.,
\protect\hyperlink{ref-archambault2013peer}{2013}; Piwowar et al.,
\protect\hyperlink{ref-piwowar2018state}{2018})

Inspired by the Hybrid OA Dashboard (Jahn,
\protect\hyperlink{ref-jahn_2017_hybridoadashboard}{2017}) we applied
licensing information detailing the legally binding access state
supplied by publishers to the publisher association Crossref to identify
OA publications. We determined the OA status of all publications
retrieved from Web of Science (henceforth WOS) and Scopus in-house
databases of 2017 by confronting them to two sources of OA information,
i.e., Unpaywall and Crossref. In Section \protect\hyperlink{data}{2}, we
present our data and methods. In Section \protect\hyperlink{results}{3}
we present our findings, while we discuss our main results in Section
\protect\hyperlink{conclusions}{4}.

\hypertarget{data}{\section{Data \& Method}\label{data}}

We queried all publications from Scopus and WOS in in-house databases of
2017. Data included article's unique ID from database and DOI. We
matched those DOIs with Unpaywall database from April
18\textsuperscript{th} 2018 to determine the OA status for each single
publication. In parallel, we matched those DOIs with Crossref data
(using snapshot of the data from April 2018 based on plus service
described in Crossref
(\protect\hyperlink{ref-crossref_2018_website}{2018})) and retrieved the
available information on the licences of publications\footnote{It is
  neccessary to note that our effort to send large number of requests to
  Crossref API (even while using plus service and through both rcrossref
  package in R and more fine-grained httr requests directly to Crossref
  API) faced timeout and response time errors and alternatively we chose
  to use the in-house snapshot of the Crossref data to circumvent the
  above error. This meant parsing large corpus of JSON files which can
  be time consuming depending on the goals of the analysis. Any effort
  on automating the proposed OA identification procedure needs to
  overcome the technical issues like this.}.

Additionally, we used the journals' ISSNs provided by Wohlgemuth,
Rimmert, \& Winterhager
(\protect\hyperlink{ref-wohlgemuth2016issn}{2016}) (and the updated
version in Rimmert, Bruns, Lenke, \& Taubert
(\protect\hyperlink{ref-rimmert2017issn}{2017})) to identify Gold OA
publications. They use different known OA indexes (e.g., DOAJ\footnote{\url{https://doaj.org/}}
(Directory of Open Access Journals) and ROAD\footnote{\url{https://road.issn.org/}}
(Directory of Open Access scholarly Resources) and determine if the
respective ISSN is listed in those databases. They differentiate between
\emph{ISSN} and \emph{ISSNL} which is more fine-grained by adding a
specific ISSN to some special issues. We tried both ISSN and ISSNL,
sicne the latter had higher matching records, therefore in our analysis
presented in the \protect\hyperlink{results}{Results} section we use the
\emph{ISSNL}.

It is necessary to note that some publications had multiple licence URLs
in Crossref database, we followed a procedure with four steps to ensure
using only \textbf{one licence per publication} (see Table
\ref{tab:description-number-of-licences-per-DOI-crossref} for the
frequencies of these publications):

\begin{enumerate}
\def\labelenumi{\arabic{enumi}.}
\tightlist
\item
  If a publication had only one record in Crossref database, whether it
  had an \emph{OA}, \emph{non-OA}, \emph{unclear} licence or \emph{no
  licence information (i.e.~NA)}, we used this status and categorized
  the publication as a unique one.
\item
  If a publication had multiple \emph{OA} licence URLs, we removed the
  duplicates and categorized it as \emph{OA}.
\item
  If a publication had a mixture of \emph{OA} and \emph{non-OA} licence
  URLs, we removed the duplicates and categorized it as \emph{OA}.
\item
  If a publication had multiple \emph{non-OA} licence URLs, we removed
  the duplicates and categorized it as \emph{non-OA}.
\end{enumerate}

A research assistant controlled the unique licences (a total of 56) we
extracted from Crossref with available information online to categorize
them as \emph{OA} and \emph{non-OA}. We used this categorization in
parallel to established OA identification procedures (e.g., searching
for journal's ISSN in DOAJ and ROAD in Gold OA identification) to ensure
a higher level of robustness in our results.

In \textbf{OA Identification process} and in order to determine if a
publication was OA or not, we applied a multi-category view separating
Gold, Hidden Gold, Hybrid and Delayed OA, while doing so, we reached a
new category of \emph{Probable Hybrid OA}. Our investigation strategy
for each category was as follows:

\begin{itemize}
\tightlist
\item
  \textbf{Gold OA}: As described earlier, we used the ISSNs provided by
  Rimmert et al. (\protect\hyperlink{ref-rimmert2017issn}{2017}) to
  determine Gold OA. We matched the respective ISSN (from both WOS and
  Scopus) with DOAJ and ROAD. If the respective ISSN was listed in one
  of those directories, the publication is categorized as \emph{Gold
  OA}. We confronted Gold OA from DOAJ and ROAD with our research
  assistant's categorization of Crossref licences after the manual check
  of unique licence URLs.
\item
  \textbf{Hidden Gold OA}: we used metadata from WOS and Scopus to
  determine the journal issue and looking at the licences of all
  publications in a single issue, if all publications had \emph{OA}
  licences, but the ISSN was not indexed in DOAJ or ROAD we categorized
  it as \emph{Hidden Gold OA}.
\item
  \textbf{Hybrid OA}: If an issue had at least one \emph{non-OA}
  publication while having one or more \emph{OA} publications, we
  categorized the OA publications as \emph{Hybrid OA}.
\item
  \textbf{Probable Hybrid OA}: If an issue did not have a \emph{non-OA}
  publication while having one or more \emph{OA} publications and some
  publications in the issue didn't have licence information, we
  categorized them as \emph{Probable Hybrid OA}.
\item
  \textbf{Delayed OA}: In all of the above cases, we looked into delays
  based on Crossref metadata (a difference in terms of days from day of
  publication and the date licence was assigned to the publication as
  described in CrossRef-API
  (\protect\hyperlink{ref-crossreff_api_2019_website}{2019}), this is
  the time period known as \emph{embargo time}) to determine if they
  were \emph{Delayed}, therefore each of the above categories were split
  to two groups, \emph{delayed} and \emph{not-delayed}. If a publication
  had multiple licence URLs on Crossref, we controlled their respective
  delay times, if any of those were \emph{not-delayed} we categorized
  the publication as such, while if any of the licences were
  \emph{delayed}, the publication is identified as a \emph{delayed} one.
\item
  \textbf{Closed Access}: Strictly speaking, if the number of
  publications in an issue was equal to the number of \emph{non-OA}
  publications and the ISSN was not indexed in DOAJ or ROAD, we
  categorized them as \emph{Closed Access}.
\item
  \textbf{NA (Not available)}: A publication that was not fitting in any
  of the above categories or did not have a licence URL to determine its
  condition was categorised as \emph{NA}. Number of NAs are higher than
  \emph{Closed Access} publications, since we aimed to keep the
  definitions as strict as possible.
\end{itemize}

\hypertarget{results}{\section{Results}\label{results}}

We present the results in two main sections, one regarding
\emph{Unpaywall} and the other on licences extracted from
\emph{Crossref}. We then present the \emph{comparison between Unpaywall
and Crossref} and the results of our \emph{manual checks on random
samples} for robustness of the results.

Table \ref{tab:description-papers-wos-and-scopus} shows the number of
\emph{articles} and \emph{review papers} from WOS and Scopus with an
equivalent record in Unpaywall database. It presents also the total
number of \emph{articles} and \emph{review papers} in WOS/Scopus to
provide a baseline for comparison. Unpaywall has higher than 50\%
coverage in both cases while coverage of WOS is slightly higher (can be
due to different indexing philosophy or DOIs completeness). In the
following tables (in \emph{Unpaywall} results), publications are limited
to only \emph{articles} and \emph{review papers} published in 2000-2017.

\rowcolors{2}{gray!6}{white}

\begin{table}[t]

\caption{\label{tab:description-papers-wos-and-scopus}All publications from WOS (2000-2017) and Scopus (2000-2017) that have an equivalent record in Unpaywall database (joined by DOIs)}
\centering
\fontsize{7}{9}\selectfont
\begin{tabular}{l|r|l}
\hiderowcolors
\hline
Data Source & Frequency & Percent\\
\hline
\showrowcolors
WOS (matched Unpaywall Only articles \& reviews) & 11,661,206 & 57.5\%\\
\hline
WOS (Only articles \& reviews 2000-2017) & 20,280,606 & -\\
\hline
Scopus (matched Unpaywall Only articles \& reviews) & 14,188,983 & 53.48\%\\
\hline
Scopus (Only articles \& reviews 2000-2017) & 26,532,295 & -\\
\hline
\end{tabular}
\end{table}

\rowcolors{2}{white}{white}

\begin{figure}

{\centering \includegraphics[width=2.67in]{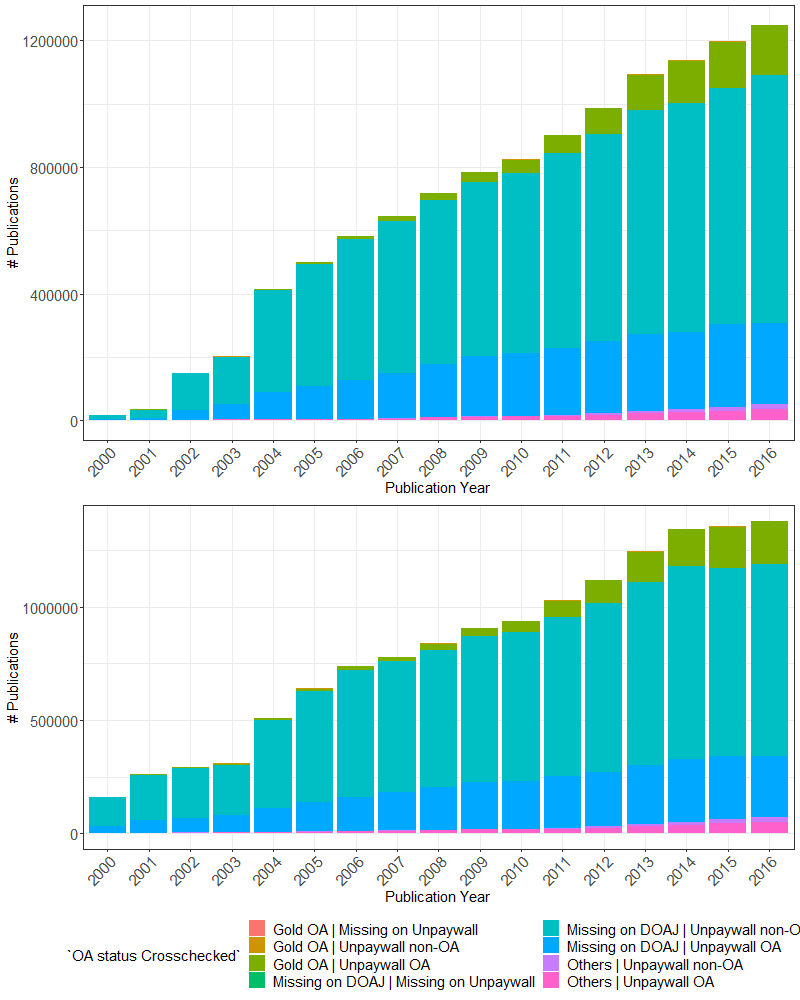} 

}

\caption{Publications indexed in WOS (top) and Scopus (bottom) matched with Unpaywall database and crosschecked the ISSNs with DOAJ (Gold OA) (X-axis denotes the years, Y-axis denotes the number of publications in each year)}\label{fig:oa-status-comparison-unpaywall-doaj-scopus-plot}
\end{figure}

Figure \ref{fig:oa-status-comparison-unpaywall-doaj-scopus-plot}
presents the distribution of journals and publications indexed in WOS
(top) and Scopus (bottom) matched with Unpaywall database and
crosschecked the ISSNs with DOAJ. \emph{Missing on DOAJ} in these
Figures refer to those journals whose ISSN was missing from Rimmert et
al. (\protect\hyperlink{ref-rimmert2017issn}{2017}) data, therefore we
could not check if the ISSN is listed in DOAJ or not while \emph{Others}
means the ISSN was existing in Rimmert et al.
(\protect\hyperlink{ref-rimmert2017issn}{2017}) but it was not listed as
\emph{OA} in DOAJ. Share of pubilcations which don't have a matching
ISSN in DOAJ (meaning they are not Gold OA) and are identified as OA in
Unpaywall is interesting on both Figures (designated with ``Missing on
DOAJ \textbar{} Unpaywall OA'' as label). They could be other OA types
(green, hybrid, hidden gold).

We matched publications to Crossref data from April 2018 and found 56
distinct licence types for all of the publications. Table
\ref{tab:description-number-of-licences-per-DOI-crossref} presents a
descriptive view on whether publications have licence information
recorded in Crossref. It shows that about 50\% of publications from WOS
or Scopus with a matching DOI indexed in Crossref do not have a licence
URL. Some of the publications had more than one licence information in
Crossref (as an example, the number of DOIs that each have 6 licence
records on Crossref are 7). In case of multiple licences, if a
publication had at least one OA licence, we categorized it as \emph{OA}.

\rowcolors{2}{gray!6}{white}

\begin{table}[t]

\caption{\label{tab:description-number-of-licences-per-DOI-crossref}Number of licences per DOI found in Crossref}
\centering
\fontsize{7}{9}\selectfont
\begin{tabular}{r|r|r}
\hiderowcolors
\hline
Number of licences per DOI & Frequency of DOIs & Percent\\
\hline
\showrowcolors
0 & 9,892,208 & 51.41\\
\hline
1 & 8,520,158 & 44.28\\
\hline
2 & 824,975 & 4.29\\
\hline
3 & 5,770 & 0.03\\
\hline
5 & 25 & 0.00\\
\hline
6 & 7 & 0.00\\
\hline
\end{tabular}
\end{table}

\rowcolors{2}{white}{white}

\begin{figure}

{\centering \includegraphics[width=2.67in]{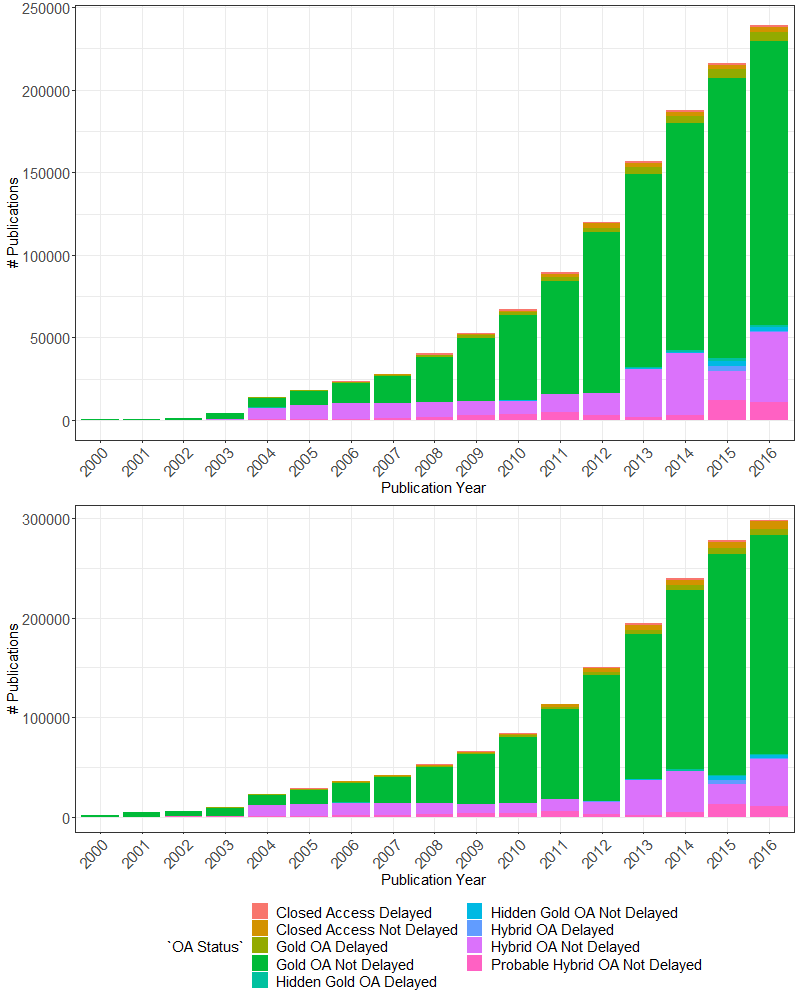} 

}

\caption{Comparison of OA publications 2000-2017 (WOS (top) and Scopus (bottom) data matched with Crossref) (X-axis denotes the years, Y-axis denotes the number of publications in each year)}\label{fig:plot-hidden-gold-oa-scopus}
\end{figure}

Figure \ref{fig:plot-hidden-gold-oa-scopus} present the \emph{Gold},
\emph{Hidden Gold}, \emph{Hybrid} and \emph{Delayed OA} status of the
publications from WOS (top) and Scopus (bottom), which is presented as
trends over the years. We limited the years to 2000-2017 to show the
most recent trends. To make these Figures more readable, we removed
\emph{NA} (those without a matching DOI or without a licence information
on Crossref).

Tables \ref{tab:comparison-unpaywall-crossref-wos} and
\ref{tab:comparison-unpaywall-crossref-scopus} present the \emph{OA
status} comparison between Unpaywall and Crossref in WOS and Scopus
publications, respectively. Note, Crossref OA status in the Tables is
the categorization we developed using respective licence URLs. We double
checked the contradictory cases and improved our while-list of OA
licences, while some of the contradictions still remain (e.g., Unpaywall
declares those publications as OA while they are closed access or vice
versa, in case of licences on Crossref that are open access while the
publication is declared as non-OA on Unpaywall). Overall contradictory
cases amount to 27.95\% in WOS and 27.57\% in Scopus which might partly
be explained by the wider scope of Unpaywall including also green OA
publications that might not be identified via license information only.

\rowcolors{2}{gray!6}{white}

\begin{table}[t]

\caption{\label{tab:comparison-unpaywall-crossref-wos}OA status comparison between Unpaywall and Crossref in WOS publications}
\centering
\fontsize{7}{9}\selectfont
\begin{tabular}{l|l|r|r}
\hiderowcolors
\hline
Crossref OA Status & Unpaywall OA Status & Frequency & Percent\\
\hline
\showrowcolors
Closed Access & Closed Access & 4,452,185 & 38.18\\
\hline
NA & Closed Access & 3,512,794 & 30.12\\
\hline
NA & Open Access & 1,770,612 & 15.18\\
\hline
Closed Access & Open Access & 1,363,525 & 11.69\\
\hline
Open Access & Open Access & 435,516 & 3.73\\
\hline
Open Access & Closed Access & 126,354 & 1.08\\
\hline
Closed Access & NA & 26 & 0.00\\
\hline
NA & NA & 19 & 0.00\\
\hline
\end{tabular}
\end{table}

\rowcolors{2}{white}{white}

\rowcolors{2}{gray!6}{white}

\begin{table}[t]

\caption{\label{tab:comparison-unpaywall-crossref-scopus}OA status comparison between Unpaywall and Crossref in Scopus publications}
\centering
\fontsize{7}{9}\selectfont
\begin{tabular}{l|l|r|r}
\hiderowcolors
\hline
Crossref OA Status & Unpaywall OA Status & Frequency & Percent\\
\hline
\showrowcolors
Closed Access & Closed Access & 5,138,444 & 36.21\\
\hline
NA & Closed Access & 4,635,801 & 32.67\\
\hline
NA & Open Access & 2,201,936 & 15.52\\
\hline
Closed Access & Open Access & 1,549,902 & 10.92\\
\hline
Open Access & Open Access & 502,510 & 3.54\\
\hline
Open Access & Closed Access & 160,132 & 1.13\\
\hline
NA & NA & 15 & 0.00\\
\hline
Open Access & NA & 4 & 0.00\\
\hline
Closed Access & NA & 1 & 0.00\\
\hline
\end{tabular}
\end{table}

\rowcolors{2}{white}{white}

Tables \ref{tab:random-sample-oa-check-wos} and
\ref{tab:random-sample-oa-check-scopus} present the result of our
research assistant's manual check for accessibility to article's PDF
file from publishers websites compared to the respecitve licence in
Crossref and the OA status we manually assignded to those URLs in
contrast to OA status from Unpaywall. It is interesting to see there are
publications defined as \emph{Non-OA} while their PDF is accessible from
the publisher (14.42\% in WOS and 14.98\% in Scopus) or vice versa, OA
publications (based on either Unpaywall, Crossref or both) that are not
accessible online (17.57\% in WOS and 16.74\% in Scopus). Note also the
contradictory cases between Crossref and Unpaywall, where metadata from
one shows \emph{OA} and the other \emph{Closed}, which requires further
probes (22.98\% in WOS and 22.91\% in Scopus, these percentages are
quite close to contradictions observed in the overall sample presented
in Tables \ref{tab:comparison-unpaywall-crossref-wos} and
\ref{tab:comparison-unpaywall-crossref-scopus}). Our effort to
complement these databases proves that none of them could be used in
isolation. We aim to follow-up and use PDF URLs provided by Unpaywall in
large scale to control the ratio of publications which can be accessed.

\rowcolors{2}{gray!6}{white}

\begin{table}[t]

\caption{\label{tab:random-sample-oa-check-wos}Random sample OA status check on publications from WOS}
\centering
\fontsize{7}{9}\selectfont
\begin{tabular}{l|l|l|r|r}
\hiderowcolors
\hline
PDF Manually accessible? & Licence status & Pub OA? & Frequency & Percent\\
\hline
\showrowcolors
PDF Accessible & Open Access & Unpaywall OA & 104 & 46.85\\
\hline
No Access to PDF & Closed Access & Unpaywall non-OA & 44 & 19.82\\
\hline
No Access to PDF & Open Access & Unpaywall non-OA & 18 & 8.11\\
\hline
No Access to PDF & Closed Access & Unpaywall OA & 16 & 7.21\\
\hline
PDF Accessible & Closed Access & Unpaywall OA & 16 & 7.21\\
\hline
PDF Accessible & Closed Access & Unpaywall non-OA & 14 & 6.31\\
\hline
No Access to PDF & Open Access & Unpaywall OA & 5 & 2.25\\
\hline
NA & Closed Access & Unpaywall non-OA & 1 & 0.45\\
\hline
No Access to PDF & Closed Access & Missing on Unpaywall & 1 & 0.45\\
\hline
PDF Accessible & NA & Unpaywall non-OA & 1 & 0.45\\
\hline
PDF Accessible & Open Access & Unpaywall non-OA & 1 & 0.45\\
\hline
PDF Accessible & NA & Unpaywall OA & 1 & 0.45\\
\hline
\end{tabular}
\end{table}

\rowcolors{2}{white}{white}

\rowcolors{2}{gray!6}{white}

\begin{table}[t]

\caption{\label{tab:random-sample-oa-check-scopus}Random sample OA status check on publications from Scopus}
\centering
\fontsize{7}{9}\selectfont
\begin{tabular}{l|l|l|r|r}
\hiderowcolors
\hline
PDF Manually accessible? & Licence status & Pub OA? & Frequency & Percent\\
\hline
\showrowcolors
PDF Accessible & Open Access & Unpaywall OA & 105 & 46.26\\
\hline
No Access to PDF & Closed Access & Unpaywall non-OA & 48 & 21.15\\
\hline
No Access to PDF & Closed Access & Unpaywall OA & 17 & 7.49\\
\hline
PDF Accessible & Closed Access & Unpaywall OA & 17 & 7.49\\
\hline
No Access to PDF & Open Access & Unpaywall non-OA & 17 & 7.49\\
\hline
PDF Accessible & Closed Access & Unpaywall non-OA & 14 & 6.17\\
\hline
No Access to PDF & Open Access & Unpaywall OA & 4 & 1.76\\
\hline
NA & Closed Access & Unpaywall non-OA & 1 & 0.44\\
\hline
PDF Accessible & NA & Unpaywall non-OA & 1 & 0.44\\
\hline
PDF Accessible & Open Access & Unpaywall non-OA & 1 & 0.44\\
\hline
PDF Accessible & NA & Missing on Unpaywall & 1 & 0.44\\
\hline
PDF Accessible & NA & Unpaywall OA & 1 & 0.44\\
\hline
\end{tabular}
\end{table}

\rowcolors{2}{white}{white}

\hypertarget{conclusions}{\section{Conclusions}\label{conclusions}}

It is clear that publishing as OA is on the rise in recent years. This
trend is observed similarly in WOS and Scopus (while Scopus has higher
raw publication counts but trends are identical) and based on OA
identification stemming from both Unpaywall and Crossref. But still the
majority of publications are closed access. We observed that despite the
high coverage of Unpaywall (higher than 50\% of \emph{articles} and
\emph{reviews} in both WOS and Scopus), it doesn't provide enough
metadata (as of April 2018) for OA categorization thus could be limiting
for large scale OA monitoring in the leading bibliometric databases.
Licence information from Crossref is more detailed and it gives a good
possibility to complement Unpaywall metadata. Although we overcame the
downsides by complementing these databases, we still found further
contradictions between them with manual random checks. Some publications
were OA (based on their licences or Unpaywall status) while their PDF
files were \emph{not accessible} through publishers' websites. Some
publications were closed access, while their PDF files were
\emph{accessible}. We found that the issue of multiple records for some
publications or multiple licence information is something that needs to
be seriously considered in OA monitoring. While we tried to test
different scenarios in OA identification, still there are publications
that won't fit into any of the scenarios and we had to categorize them
as \emph{NA} (since we wanted to keep the \emph{Closed Access}
definition as strict as possible), these are the publications that need
to be further studied and usually the metadata of the OA databases are
lacking for them. We propose OA monitoring activities to try to benefit
from our approach in compelemting the metadata from OA databases,
i.e.~Unpaywall and Crossref, while noting that there are contradictions
between these sources. Our effort to complement these databases proves
that none of them could be used in isolation.

\section*{References}\label{references}
\addcontentsline{toc}{section}{References}

\hypertarget{refs}{}
\hypertarget{ref-archambault2013peer}{}
Archambault, E., Amyot, D., Deschamps, P., Nicol, A., Rebout, L., \&
Roberge, G. (2013). Peer-reviewed papers at the european and world
levels---2004-2011. \emph{Info@ Science}, \emph{1}, 495--6505.

\hypertarget{ref-crossref_2018_website}{}
Crossref. (2018, October). Crossref. Retrieved from
\url{https://www.crossref.org/}

\hypertarget{ref-crossreff_api_2019_website}{}
CrossRef-API. (2019, October). CrossRef api. Retrieved from
\url{https://github.com/CrossRef/rest-api-doc\#filter-names}

\hypertarget{ref-jahn_2017_hybridoadashboard}{}
Jahn. (2017, January). About the hybrid oa dashboard. Retrieved from
\url{https://subugoe.github.io/hybrid_oa_dashboard/about.html}

\hypertarget{ref-piwowar2018state}{}
Piwowar, H., Priem, J., Larivière, V., Alperin, J. P., Matthias, L.,
Norlander, B., \ldots{} Haustein, S. (2018). The state of oa: A
large-scale analysis of the prevalence and impact of open access
articles. \emph{PeerJ}, \emph{6}, e4375.

\hypertarget{ref-rimmert2017issn}{}
Rimmert, C., Bruns, A., Lenke, C., \& Taubert, N. C. (2017).
ISSN-matching of gold oa journals (issn-gold-oa) 2.0.

\hypertarget{ref-wohlgemuth2016issn}{}
Wohlgemuth, M., Rimmert, C., \& Winterhager, M. (2016). ISSN-matching of
gold oa journals (issn-gold-oa).

\end{document}